# *On the choice of coarse variables for dynamics*

Amit Acharya
Civil and Environmental Engineering, Carnegie Mellon University, Pittsburgh, PA 15213
Ph. – (412) 268 4566, Fax – (412) 268 7813, email- acharyaamit@cmu.edu

Abstract

Two ideas for the choice of an adequate set of coarse variables allowing approximate autonomous dynamics for practical applications are presented. The coarse variables are meant to represent averaged behavior of a fine-scale autonomous dynamics.

## 1. Introduction

The goal of this paper is to lay out some ideas for the further development of an invariant-manifold-theory inspired computational approach to the problem of coarse-graining an autonomous system of ODE (fine system). Coarse variables are introduced as either functions of the fine state or time-averages of functions of the fine state. The objective is to come up with a closed theory of evolution for the coarse variables. In our past work [1, 2, 3] that we have called the method of Parametrized Locally Invariant Manifolds (PLIM), we have shown that this goal can be achieved in the context of hard nonlinear problems involving dissipation and/or oscillatory response. Strictly speaking, the achievement is, however, only partial in the sense that one requires knowledge of fine initial conditions to ensure a correct coarse-grained response even though the developed coarse equations are posed purely in terms of the coarse variables. This realization is intimately tied to the understanding of the emergence of memory effects in coarse response of an autonomous fine theory, a feature that can also be interpreted as stochastic effects in coarse response. In this paper, we examine two methods for the selection of a small number of coarse variables designed to allow for an autonomous coarse response, thus allowing unambiguous initialization of the coarse theory with information only on the coarse state.

PLIM is an algorithm for developing approximate, but micro-dynamics-consistent, equations of evolution for user-defined coarse variables. The broad idea here is to calculate parametrizations of an appropriate *collection* of locally invariant manifolds of the fine dynamics *a priori*, with the coordinates being the coarse variables (observables). With a database devised to store this information, it becomes possible to define a closed dynamics for the coarse variables.

PLIM as well as the Mori-Zwanzig Projection Operator Technique imply that if coarse variables are chosen 'arbitrarily', then more variables will, in general, be required to have an autonomous coarse theory that can be initialized unambiguously. Here, we propose two possible strategies for making a choice of such coarse variables. Connections of our work with, and a detailed review of, other multiscale strategies are provided in [1,2,3]. In particular, we note the work of Kevrekidis and co-workers [10,11,12] where the emphasis is not on deriving the form of the coarse equations at all but nevertheless make consistent predictions of coarse evolution based on a carefully crafted strategy for utilizing short bursts of microscopic simulations.

The *main point* of the paper is how to start from a fine dynamics with some idea of what time-averaged coarse variables one might be interested in and proceed to augment this problem in a well-defined manner so that coarse response can indeed be computed. In a rough sense, it is specifically designed to deal with problems where the 'as-received' fine scale problem does not readily have an obvious slow macroscopic dynamics associated with it. The procedure tries to augment the problem definition so that an appropriate macroscopic dynamics becomes associated with the augmented microscopic problem.

This work is mathematically formal.

## 2. Background

The autonomous fine dynamics is defined as

$$\frac{df}{dt}(t) = H(f(t)) \tag{1}$$
$$f(0) = f_*.$$

$f$ is an $N$-dimensional vector of fine degrees of freedom and $H$ is a generally nonlinear function of fine states, denoted as the vector field of the fine dynamical system. Equation $(1)_2$ represents the specification of initial conditions. $N$ can be large in principle, and the function $H$ rapidly oscillating.

Let $\Lambda$ be a user-specified function of the fine states producing vectors with $m$ components whose time averages over intervals of period $\tau$ can be measured in principle and are of physical interest. These time averages are considered as some of the coarse variables of interest. Let us also define the remaining list of 'instantaneous' coarse variables $p$ with $n$ components through the relationship

$$p = \Pi(f), \qquad (2)$$

where $\Pi$ is user-defined. Often, such variables may be required to incorporate external driving influences (e.g. loadings) on an assembly whose time-averaged behavior needs to be explored.

Given the fixed time interval $\tau$ characterizing the resolution of coarse measurements in time, a coarse trajectory corresponding to each fine trajectory $f(\cdot)$ is defined as the following pair of functions of time:

$$c(t) = \frac{1}{\tau} \int_{t}^{t+\tau} \Lambda(f(s)) ds,$$
$$p(t) = \Pi(f(t)). \qquad (3)$$

Roughly speaking, it is a closed statement of evolution for the pair $(c, p)$ that we seek. The statement is unambiguous only after we specify what sort of initial conditions we may want to prescribe. For the purpose of this section we assume that fine initial conditions are known with certainty. Then, the goal is to develop a closed evolution equation for $(c, p)$, i.e. an equation that can be used for evolving $(c, p)$ without concurrently evolving (1), corresponding to fine trajectories out of a prescribed set of fine initial conditions.

Clearly,

$$\frac{dc}{dt}(t) = \frac{1}{\tau}\left[\Lambda(f(t+\tau)) - \Lambda(f(t))\right]$$
$$\frac{dp}{dt}(t) = \sum_{J=1}^{N} \frac{\partial \Pi}{\partial f^J}(f(t)) H^J(f(t)) \qquad (4)$$

If we now introduce a forward trajectory $f_f(\cdot)$ corresponding to a trajectory $f(\cdot)$ as

$$f_f(t) := f(t+\tau), \qquad (5)$$

then

$$\frac{df_f}{dt}(t) = H(f_f(t)). \qquad (6)$$

Also, given an initial state $f_*$ we denote by $f_{**}$ the state defined as the solution of (1) evaluated at time $\tau$. With these definitions in hand, we augment the fine dynamics (1) to

$$\frac{df_f}{dt}(t) = H(f_f(t))$$

$$\frac{df}{dt}(t) = H(f(t)) \quad (7)$$

$$f_f(0) = f_{**}$$

$$f(0) = f_*$$

and apply invariant manifold techniques to (7). In detail, on an $m+n$- dimensional coarse phase space whose generic element we denote as $(c, p)$, we seek functions $G_f$ and $G$ that satisfy the first-order, quasilinear partial differential equations

$$\left.\begin{aligned}\sum_{k=1}^{m}\frac{\partial G_f^I}{\partial c^k}\left\{\frac{1}{\tau}\left[\varLambda^k(G_f) - \varLambda^k(G)\right]\right\} + \sum_{l=1}^{n}\sum_{K=1}^{N}\frac{\partial G_f^I}{\partial p^l}\frac{\partial \varPi^l}{\partial f^K}(G)H^K(G) = H^I(G_f) \\ \sum_{k=1}^{m}\frac{\partial G^I}{\partial c^k}\left\{\frac{1}{\tau}\left[\varLambda^k(G_f) - \varLambda^k(G)\right]\right\} + \sum_{l=1}^{n}\sum_{K=1}^{N}\frac{\partial G_f^I}{\partial p^l}\frac{\partial \varPi^l}{\partial f^K}(G)H^K(G) = H^I(G)\end{aligned}\right\} \quad I = 1 \text{ to } N \quad (8)$$

at least locally in $(c, p)$-space. Assuming that we have such a pair of functions over the domain containing the point

$$c_* := c(0)$$
$$p_* := p(0) \quad (9)$$

defined from (1) and (3) which, moreover, satisfies the conditions

$$G_f(c_*, p_*) = f_{**}$$
$$G(c_*, p_*) = f_*, \quad (10)$$

it is easy to see that a local-in-time fine trajectory defined by

$$\Gamma_f(t) := G_f(c(t), p(t))$$
$$\Gamma(t) := G(c(t), p(t)) \quad (11)$$

through the coarse local trajectory satisfying

$$\frac{dc}{dt} = \frac{1}{\tau}\left[\varLambda(G_f(c, p)) - \varLambda(G(c, p))\right]$$

$$\frac{dp}{dt} = \sum_{J=1}^{N}\frac{\partial \varPi}{\partial f^J}(G(c, p))H^J(G(c, p)) \quad (12)$$

$$c(0) = c_*$$
$$p(0) = p_*$$

is the solution of (7) (locally). A solution pair $(G_f, G)$ of (8) represents a parametrization of a locally invariant manifold of the dynamics (7). By a locally invariant manifold we mean a set of points in phase space such that the vector field of (7) is tangent to the set at all points. Thus, a trajectory of (7) exits a locally invariant manifold only through the boundary of the manifold.

Also, note that if $(\widehat{G}_f, \widehat{G})$ and $(\breve{G}_f, \breve{G})$ are two solutions to (8) and (10) on an identical local domain in $(c, p)$-space containing $(c_*, p_*)$ and $(\widehat{c}(\cdot), \widehat{p}(\cdot))$ and $(\breve{c}(\cdot), \breve{p}(\cdot))$ are the corresponding coarse trajectories defined as solutions to (12), then local uniqueness of solutions to (7) implies

$$\widehat{G}_f(\widehat{c}(t), \widehat{p}(t)) =: \widehat{\Gamma}_f(t) = \breve{\Gamma}_f(t) := \breve{G}_f(\breve{c}(t), \breve{p}(t))$$
$$\widehat{G}(\widehat{c}(t), \widehat{p}(t)) =: \widehat{\Gamma}(t) = \breve{\Gamma}(t) := \breve{G}(\breve{c}(t), \breve{p}(t)). \tag{13}$$

Thus,

$$\frac{d\widehat{c}}{dt}(t) = \frac{d\breve{c}}{dt}(t) \quad ; \quad \frac{d\widehat{p}}{dt}(t) = \frac{d\breve{p}}{dt}(t) \quad \text{locally in time},$$
$$\widehat{c}(0) = \breve{c}(0) \quad ; \quad \widehat{p}(0) = \breve{p}(0) \tag{14}$$

from (12), and assuming $(\widehat{c}, \widehat{p})$ and $(\breve{c}, \breve{p})$ are continuous, $(\widehat{c}, \widehat{p}) \equiv (\breve{c}, \breve{p})$, locally.

Hence, given *any* pair of mappings $G_f, G$ satisfying (8) *and* (10) on a domain containing $(c_*, p_*)$, we consider (12) as the *consistent, closed* theory for the evolution of the coarse variables $c$. Obstruction to the construction of solutions to (8) is explored in [1], providing one reason for seeking multiple local solutions as implemented in [2]. This paper seeks to determine coarse functions such that multiple local solutions are not required, as far as possible.

Notice that if equations (4)$_1$, (5), (7) are viewed as a system, then this system has a singular perturbation structure for $\tau$ large. We are interested in the evolution of the '$c$' variables which are coarse/macroscopic variables. Section 3 briefly outlines a concrete and systematic procedure of how to append this system with more memory variables (in mechanics parlance, internal variables) so as to obtain an unambiguously initializable coarse dynamics.

As explained in [1,2,3], an arbitrary choice of coarse variables (12) will not, in general, result in unique evolution of the coarse state out of a specified coarse initial condition. As well, conservation properties of the fine system are not expected to be preserved in the behavior of the time averages of fine variables. Thus, fixing attention on a fixed set of arbitrarily chosen coarse variables would imply what would seem like stochastic coarse response with dissipation. This is also the content of the main result of the Mori-Zwanzig projection Operator Technique within which Langevin dynamics can be derived. In this work, we propose a different strategy – we would like to start with a particular physically motivated set of coarse variables, but then would like to augment this set with more appropriately chosen variables so that the augmented set displays autonomous response. These extra variables effectively are memory (delay) variables corresponding to the original set of coarse variables. In the next section, we outline a procedure for the selection of such extra variables and then the use of the PLIM methodology to set up the autonomous coarse response.

**3. Variables for autonomous coarse response: The Delay Reconstruction Technique**

The main conceptual ingredient of this technique is Takens's embedding theorem [6]:

> *Theorem:* Let $M$ be a compact manifold of dimension $m$. For pairs $(\varphi, y)$, $\varphi: M \to M$ a smooth diffeomorphism and $y: M \to R$ (reals) a smooth function, it is a generic property that the map $\Phi_{(\varphi, y)}: M \to R^{2m+1}$, defined by
>
> $$\Phi_{(\varphi, y)}(x) = (y(x), y \circ \varphi(x), \cdots, y \circ \varphi^{2m}(x))$$

is an embedding; by 'smooth' we mean at least $C^2$.

Practically, Takens's theorems suggest, and were motivated by the idea, that a single measurable signal (the function $y$) of a complicated, possibly high-dimensional, dynamics (the mapping $\varphi$), can in principle reveal all qualitative features of the underlying dynamics through the study of the delay-reconstruction map.

Thus the delay reconstruction technique has been used by workers to make statements about qualitative features of dynamics. Of particular interest is the work in [4] and [5] where an algorithm for a systematic unfolding of delay-reconstructed trajectories is introduced, corresponding to an autonomous original dynamics. Essentially, starting from a one-component delay reconstruction of a trajectory of the original dynamics, more delay components are added till the point where the trajectory in delay-reconstruction space has no self-intersections. The number of components required is then declared to be representative of the number $2m$ of the theorem. Ding et al. [7] show that provided the function $y$ satisfies certain smoothness assumptions, the correlation dimension of the delay-reconstructed signal with progressively more components hits a plateau when it becomes just greater than the correlation dimension of the attractor of the original dynamics.

On the other hand, physical intuition suggests that any kind of measuring device acts as a filter that cannot measure variations below its resolution. Thus if $y$ were to represent a moving time average, then it could not possibly reveal all features of the original dynamics. In a deep physical sense, were this not the case, macroscopic physics would not be possible – time-averaged signals display much gentler and lower dimensional dynamics. It is this idea that we would now like to pursue. Suggestive practical examples of this feature of dynamics can be found in [4].

Consider Axiom A diffeomorphisms for which we have ergodicity on attractors. Consider an observable of the form

$$y'(x) = \lim_{N \to \infty} \frac{1}{N} \sum_{p=0}^{N} \varphi^p(x), \qquad (15)$$

where we assume that $\varphi$ takes values in $\mathbb{R}^k$ for some positive integer $k$. Due to ergodicity, this map has a constant value almost everywhere on a set of non-vanishing Lebesgue measure in $\mathbb{R}^k$ containing the attractor. Thus the set of points generated by the dynamics corresponding to this observable along almost all trajectories on the attractor has dimension $0$, whereas the original attractor has some finite, possibly large, dimension. If $y'$ (or each of its component functions) satisfied the smoothness hypotheses of Takens's theorem, then this would be a contradiction, as can be seen by utilizing a theorem of Eckmann and Ruelle [8] related to the determination of correlation dimension. Of course, it does not, since the value of $y'$ at a fixed point of $\varphi$ is the fixed point itself whereas for an evaluation at a slight perturbation off the fixed point, the value is the ergodic average implying that the observable is, in all likelihood, discontinuous, let alone smooth. If we step back from an infinite sum as in (15) and perform finite sums with large $N$ for realistic dynamics where the map $\varphi$ may not have enough smoothness to be a diffeomorphism (e.g. the interatomic force for the Lennard-Jones potential contains odd powers of square roots of the interatomic distance), we do not expect the above argument to change drastically, in the sense of a discontinuous function being approached as a limit of smooth functions. Therefore, time-averaged observables, reflecting real measurements, maybe expected to display lower-dimensional dynamics. Moreover, such variables may be very useful for coarse behavior.

Thus the idea with regard to definition of an autonomous coarse dynamics is to choose a set of physically motivated, time-averaged coarse variables corresponding to the original high-dimensional dynamics. An aperiodic, dense(on the attractor) original trajectory is then delay-reconstructed in terms of these variables plus added delay components, up to the point where the reconstructed trajectory has no self-intersections. The original set of coarse variables plus their delay counterparts form the augmented set of coarse variables that form an autonomous coarse dynamics. One now uses PLIM with this set of coarse variables to establish the coarse dynamics. The procedure outlined in Section 2 in dealing with one set of delay variables $\left(f_f\right)$ is easily generalized to deal with multiple (sets of) delay variables with different delays.

If indeed the delay-reconstructed trajectory has no self-intersection and the embedding dimension is small compared to the dimension of the attractor or its enveloping inertial manifold for the original dynamics, then this implies that we have a one-to-one mapping between sets of different dimension. In this connection, the work of Sauer et al. [9] should be noted, especially their filtered delay embedding theorem. Again there are strong hypotheses involved, and, interestingly, their Self-Intersection theorem does not put a lower-bound on the dimension of the self-intersection set, thus leaving a ray of hope with regard to the existence of a one-one map. Of course, it should be noted that one–to-one maps between sets of different dimensions can be continuous (in the small-to-large dimension direction) but nowhere differentiable, but this may be acceptable in the PLIM approximation methodology in approaching such functions as limits of piecewise-smooth continuous functions.

As an example of the application of these ideas, one may consider the Frenkel-Kontorova model of a chain of atoms interacting through linear springs as well as with a nonlinear substrate potential. The chain may be assumed fixed at one end and a load applied at the other. Of interest is the time averaged stress strain (end-load-displacement) curve of this 1-d assembly. PLIM is applied to this problem setup in [2]. However, strictly speaking (and as mentioned in [2]), the problem there is solved only for coarse evolution corresponding to the fine trajectory starting from the stress free initial state. With the developments suggested in this paper, it may be hoped that the two average stress and strain coarse variables can be systematically augmented with more memory variables such that coarse evolution based on the developed theory is provably representative (at least in a formal sense) of coarse evolution corresponding to a large class of fine trajectories.

*Remark*: We note here that in the conventional applications of the delay-reconstruction technique, non-generic or non-smooth observables leading to dimension change, e.g. Broomhead et al. [13] and Pecora and Carroll [14], are considered anomalous and to be avoided. Of particular interest to this work, Broomhead et al. [13] explicitly construct an example involving a nonrecursive filter, the inverse all-pole filter, that *reduces* dimension under delay-reconstruction of a particular signal. For the question of coarse-graining/averaging, however, it seems that it is precisely such non-generic and/or non-smooth functions that should be relevant. Indeed, it makes sense that a set of coarse observables executing autonomous macroscopic dynamics is special and cannot be chosen as any generic, smooth function(s).

## 4. Variables for autonomous coarse response: Adapted projections

In this proposed approach to find coarse variables that evolve autonomously, we consider the following argument: let $\Pi$ be a scalar function on the fine phase space, representing the definition of the sought-for coarse variable. Let $f$ be a fine trajectory. Defining the coarse trajectory corresponding to $f$ by $c$, we have

$$c(t) := \Pi(f(t)) \quad \Rightarrow \quad \dot{c}(t) = \frac{\partial \Pi}{\partial f}(f(t)) H(f(t)). \tag{16}$$

Now, in general, given $\Pi$, many fine states will correspond to a single coarse state. Under the circumstances, one way to ensure an autonomous coarse response is to ensure that on any set of fine states where the evaluation of $\Pi$ agrees, so should the right-hand-side of the $c$ evolution. Mathematically, we have the following: we are interested in determining a function $\Pi$ that has the following property. Let $c$ be arbitrarily fixed. Then define

$$W_c := \{f : \Pi(f) = c\}. \tag{17}$$

Now require that $\Pi$ satisfy

$$\frac{\partial \Pi}{\partial f}(f) H(f) = A_c \text{ on } W_c, \tag{18}$$

where $A_c$ is a constant depending on $c$ but independent of $f$.

Equations (17) and (18) together imply that a level set of $\Pi(\cdot)$ should also be a level set of the function $\frac{\partial \Pi}{\partial f}(\cdot) H(\cdot)$. One way to require this is to demand that

$$\lambda \frac{\partial}{\partial f}(\Pi) = \frac{\partial}{\partial f}\left(\frac{\partial \Pi}{\partial f} H\right) \tag{19}$$

where $\lambda$ is an arbitrary scalar. Choosing it to be a derivative of an arbitrary function of a single variable, i.e. of the form $\lambda = \partial \varphi / \partial \Pi$, we have the necessary condition that

$$\frac{\partial}{\partial f}\left[\varphi(\Pi) - \frac{\partial \Pi}{\partial f} H\right] = 0. \tag{20}$$

Thus, if we now require $\Pi$ to satisfy the first-order linear PDE on the fine phase space

$$\varphi(\Pi) = \frac{\partial \Pi}{\partial f} H, \tag{21}$$

it can be show by reversing the above arguments that

$$\dot{c} = \varphi(c) \tag{22}$$

would be the correct *autonomous,* coarse evolution equation for the coarse variable defined by $\Pi$ obtained as a solution to (21), for arbitrary choices of $\varphi$ in (21). Thus an entire class of coarse variables can be defined based on the choice of $\varphi$, which is a somewhat surprising result. It may be hoped that this class contains physically meaningful coarse variable definitions. More importantly, it perhaps suggests that the choice of appropriate coarse variables cannot be left completely unconstrained, and their definition requires physical guidance.

Equation (21) is a linear first order PDE for $\Pi$. Characteristic curves for the PDE are solutions to the original set of fine system of ODE. They do not intersect (and meet only at fixed points) because of the autonomous nature and smoothness of the fine evolution. Thus shocks do not exist, and it appears reasonable to expect to numerically approximate the PDE without any further conditions.

**5. Acknowledgments:** It is a pleasure to acknowledge discussions with Luc Tartar and Noel Walkington. In particular, LT suggested the local characterization (19) and NW pointed out the backward-in-time uniqueness for smooth ODE.

# 6. References

[1] Acharya, A., Parametrized invariant manifolds: A recipe for multiscale modeling? Computer Methods in Applied Mechanics and Engineering, 194, 3067-3089, 2005.

[2] Acharya, A. and Sawant, A., On a computational approach for the approximate dynamics of averaged variables in nonlinear ODE systems: toward the derivation of constitutive laws of the rate type, Journal of the Mechanics and Physics of Solids, 54, 2183-2213.

[3] Sawant, A., Acharya, A., Model reduction via Parametrized Locally Invariant manifolds: Some Examples, Computer Methods in Applied Mechanics and Engineering, 195, 6287-6311, 2005.

[4] Abarbanel, H. D. I., Analysis of observed chaotic data, Institute for Nonlinear Science Series, Springer-Verlag.

[5] Kennel, M. B., Abarbanel, H. D. I. False neighbors and false strands: a reliable minimum embedding dimension algorithm, Physical review E, 66, 026209, 2002.

[6] Takens, F., Detecting strange attractors in turbulence, In: Dynamical Systems and Turbulence, Lecture Notes in Mathematics, 898, Warwick 1980, Ed. D. A. Rand, L. S. Young, 366-381, 1980.

[7] Ding, M., Grebogi, C., Ott, E., Sauer, T., Yorke, J. A. Estimating correlation dimension from a chaotic time series: when does plateau onset occur? Physica D, 69, 404-424, 1993.

[8] Eckmann J.-P., Ruelle, D. Ergodic theory of chaos and strange attractors, Reviews of Modern Physics, 57, 3, 617-656, 1985.

[9] Sauer, T., Yorke, J. A., Casdagli, M. Embedology, Journal of Statistical Physics, 65, 3/4, 579-616, 1991.

[10] Gear, C. W., Kevrekidis, I. G., Theodoropoulos, C., Coarse Integration/Bifurcation Analysis via Microscopic Simulators: micro-Galerkin methods, , Comp. Chem. Engng., 26, 941-963, 2002.

[11] Kevrekidis, I. G., C. W. Gear, J. M. Hyman, P. G. Kevrekidis, O. Runborg and K. Theodoropoulos, Equation-free coarse-grained multiscale computation: enabling microscopic simulators to perform system-level tasks, Comm. Math. Sciences, 1(4), 715-762, 2003.

[12] Kevrekidis, I. G., C. William Gear, G. Hummer, Equation-free: the computer-assisted analysis of complex, multiscale systems, A.I.Ch.E Journal, 50(7), 1346-1354, 2004.

[13] Broomhead, D. S., Huke, J. P., Muldoon, M. R. Linear filters and non-linear systems, Journal of the Royal Statistical Society, Ser. B, 54(2), 373-382, 1992.

[14] Pecora, L. M., Carroll, T. L. Discontinuous and non-differentiable functions and dimension increase induced by filtering chaotic data, Chaos, 6(3), 432-439, 1996.